\documentclass[twocolumn,showpacs,preprintnumbers,amsmath,amssymb]{revtex4-1} 
\usepackage{graphicx}% Include figure files
\usepackage{dcolumn}% Align table columns on decimal point
\usepackage{bm}% bold math
\usepackage{color}
\usepackage{bbold}

\def\beq{\begin{equation}} 
\def\eeq{\end{equation}} 
\def\pmb#1{\setbox0=\hbox{#1}%   
 \kern-.025em\copy0\kern-\wd0 
 \kern .05em\copy0\kern-\wd0 
 \kern-.025em\raise.0433em\box0 } 
\def\sig{\sigma}

\def\vr{\vec{r}}

\def\vx{\vec{x}}

\def\vu{\vec{u}}
\def\vv{\vec{v}}

\def\vF{\vec{F}}
\def\vN{\vec{N}}
\def\vR{\vec{R}}
\def\vLambda{\vec{\Lambda}}

\def\vDelta{\vec{\Delta}}

\def\vV{\vec{V}}

\def\vrho{\vec{\rho}}
\def\vnu{\vec{\nu}}
\def\vphi{\vec{\phi}}
\def\vom{\vec{\omega}}

\def\he{{\hat e}}

\def\halpha{{\hat \alpha}}

\begin{document} 

%\title{First-principles derivation of flow equations for dense granular fluids}
\title{Flow equations for dense granular fluids: New insight from a first-principles derivation}

\author{Moshe Schwartz}
\affiliation{Beverly and Raymond Sackler School of Physics and Astronomy, Tel Aviv University, Ramat Aviv 69934, Israel}

\author{Raphael Blumenfeld}
\email[]{rbb11@cam.ac.uk}
%\homepage[]{Your web page}
%\thanks{}
\altaffiliation{Also at: Cavendish Laboratory, JJ Thomson Avenue, Cambridge CB3 0HE, UK}
\affiliation{Imperial College London, London SW7 2AZ, UK}

\date{\today}
%\maketitle

\begin{abstract}

We present a first-principles theory for plug-free dense granular flow. This is done by coarse-graining directly the microscopic dynamics and deriving an explicit relation between the macroscopic stress and strain rate tensors. The newly derived relation not only differs significantly from that of the existing empirical models for such flows but it also provides a novel understanding of the effect of rigid-like rotational regions in the flow.

\end{abstract}


¤

\maketitle

%-----------------------Begin introduction

The significance of flow of dense granular matter to many natural and man-made phenomena, as well as to a uniquely large number of disciplines, cannot be overemphasized. The reproducible patterns of such flow on scales much larger than the typical grain size, suggests that, as in more conventional fluids, it should be possible to model this phenomenon with continuum flow equations. This has been the goal of much recent research.  
Existing first-principles derivation of flow equations from microscopic considerations are valid only for dilute fluids, being based on the theory of binary collisions between particles. This derivation breaks down for slow dense granular fluids due to the prolonged contacts between particles, which make irrelevant the concept of collisions.
Yet, since the forces between touching grains are well understood, it is plausible that continuum flow equations for this regime can be derived under appropriate coarse-graining. 
The derivation of a fundamental theory from microscopic dynamics is of major significance to the field, as it makes possible generally applied predictions. Yet, recent flow models are based on empirical stress - strain rate relations (SSRRs)\cite{MiDi,FoPo08}. Recent models use solid friction, described first by da Vinci\cite{dV}, Amontons\cite{Am1699}  and Coulomb\cite{Co1779}, as the dissipating mechanism\cite{FoPo08,Sc87,Jop06}, rather than traditional viscosity. However, these models are based on a naive expectation on the form of the SSRR that closes the flow equations.

Here, we use the same dissipative mechanism, but we aim at a first-principles flow theory for this regime and derive a macroscopic SSRR by a direct coarse-graining from the grain scale dynamics.
The relation we find not only differs significantly from the naive empirical form in the literature but also shows that the latter misses a crucial aspect - the stress dependence on local rigid-like rotational flow.

The inter-granular forces can be decomposed into normal contact forces and solid friction forces. 
The solid friction dissipates mechanical energy into intra-granular degrees of freedom (e.g. heat), in contrast to ordinary fluids, where mechanical energy of macroscopic fluid disturbances is dissipated by viscosity into fluctuations of much smaller scale.

A significant feature of dense granular fluids is their tendency to form under some conditions plug regions that move as rigid bodies within the fluid. These regions 
play an essential role in the flow and have been the subject of two recent papers\cite{BlScEd08,ScBl11}. 
To complete the description, we focus here on  deriving the SSRR in plug-free regions of dense granular flow.  

Our approach is based on identifying two separate contributions to the stress tensor: $\sig^{(n)}$, resulting from normal contact forces, and $\sig^{(f)}$, resulting from friction forces. The central problem, solved here, is relating $\sig^{(f)}$, $\sig^{(n)}$ and the strain rate tensor $T$.

%-----------------------End introduction

\begin{figure} ["here"]
\includegraphics[width=8.5cm]{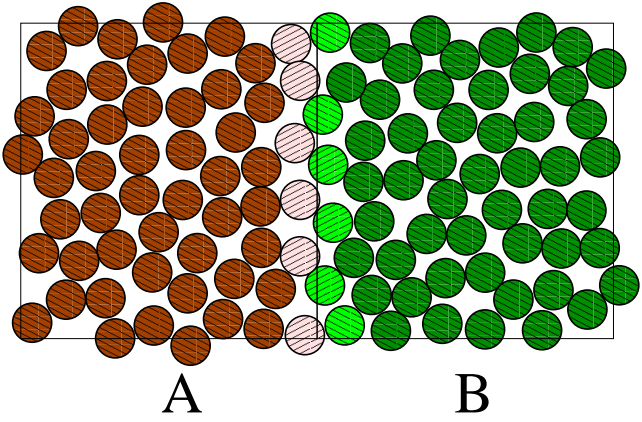}%
\caption{\label{Fig1} 
Volumes $A$ and $B$ are divided by a mathematical boundary plane. Boundary grains (lighter colours) are in contact, transmitting normal and friction forces.
}
\end{figure}

Consider a dense system of roughly spherical rigid convex grains, of typical size $d$, interacting via normal contact and tangential friction forces. Our aim is to coarse-grain these interactions to obtain the effective interaction between neighbouring volume elements of the fluid, which are large enough to contain many grains, but are much smaller than the system size. 
Two such volume elements, $V_A$ and $V_B$, are shown in figure  \ref{Fig1}. The virtual boundary plane separating these volume may cut individual grains, which are deemed to belong to $V_A$ or $V_B$, depending on the locations of their centres of mass.

We wish to obtain the net force per unit area applied by grains in $V_A$  to grains in $V_B$ near a point $\vx$ on the boundary plane. We consider separately the contributions to this force from the non-dissipative normal contact forces and from the dissipative frictional (tangential) forces, where the normal and tangential directions are with respect to the inter-granular tangent contact planes. 
Let $k$ denote a pair  of grains, $i$ and $j$, touching across the $A-B$ boundary plane and belonging to $V_A$ and $V_B$, respectively.  
Let $\vN_k$ be the normal contact force that the $A$ member of the pair applies to the $B$ member and $\vr_{ij}$ the position vector from the centre of grain $i$ to its contact with grain $j$. The solid friction force between the members of the pair depends on whether they slide relative to one another at the contact or not. The velocity of grain $i$ relative to $j$ is,

\beq
\vLambda_k \equiv \vDelta_k + \frac{1}{2}\left(\vom_i+\vom_j\right)\times\vR_k + \frac{1}{2}\left(\vom_i-\vom_j\right)\times\vrho_k
\label{kvelocity}
\eeq
where $\vDelta_k=\vv_i-\vv_j$ is the relative centre of mass velocity of the two grains, 
$\vom_i$ and $\vom_j$ are, respectively, the angular velocities of the two grains around their centres of mass, 
$\vR_k=\vr_{ij}-\vr_{ji}$ is the vector from the centre of mass of grain $i$  to the centre of mass of grain  $j$ and
$\vrho_k=\vr_{ij}+\vr_{ji}$.
Note that $\vrho_k=0$ for identical spherical grains. 

When $\vLambda_k\neq 0$ the $A$ grain applies to the $B$ grain a friction force given by the da Vinci - Amontons - Coulomb law

\beq
\vF_k = \mu_d |\vN_k| \vu_k 
\label{dVAClaw}
\eeq
where $\mu_d$ is the dynamic friction coefficient between the members of the pair and $\vu_k=\vLambda_k/|\vLambda_k|$. For simplicity, we assume the same $\mu_d$ between all rubbing particles.
At the contact we have a `rubbing condition', 

\beq
\vLambda_k\cdot \vN_k=0 
\label{RubCon}
\eeq
This is because $\vLambda_k\cdot \vN_k > 0$ corresponds to inter-penetration of the pair, which is impossible due to their rigidity, and $\vLambda_k\cdot \vN_k < 0$ corresponds to loss of contact, in which case $\vN_k$ vanishes. 
When $\vLambda_k = 0$, we only know that the inter-granular friction force satisfies $\vF_k\cdot\vN_k=0$ and $|\vF_k|\leq \mu_s |\vN_k|$, where $\mu_s (\geq \mu_d)$ is the static friction coefficient between members of the pair. In that case a further condition is required to determine $\vF_k$.

Denoting by $\vx_k$ the intersection of the vector $\vR_k$ with the boundary plane, 
the average normal force per unit area exerted by the volume element  $V_A$ on the volume element $V_B$  is 

\beq
\vnu(\vx) = \pi(\vx) \vN(\vx)
\label{NormalForceDensity}
\eeq
where $\pi(\vx)$ is the pair density per unit area at $\vx$ and $\vN(\vx)=\langle\vN_k\rangle$ is the average normal contact force applied by an $A$ member of the pair to its $B$  member near $\vx$. 
This is simply the spatial average over a circular area centered at $\vx$ and located on the plane separating $V_A$ from $V_B$. 
The circle is sufficiently small for the difference of average quantities across it  to be adequately represented by the first order Taylor series, yet sufficiently large to be traversed by a statistically meaningful number of pairs.

Similarly, the average solid friction force per unit area, exerted by the volume element  $V_A$ on $V_B$, is 

\beq
\vphi(\vx) =  \pi(\vx) \vF(\vx)
\label{FrictionForceDensity}
\eeq
where $\vF(\vx)$ is the average friction force exerted by an $A$ member of a pair to the $B$ member near $\vx$. 
To obtain the average solid friction force and express it in terms of the normal stress tensor $\sig^{(n)}$ and the strain rate  tensor $T$, we have to discuss the  coarse-grained quantities and the corresponding  fluctuations that go into the calculation of the average. 
First we separate $\vLambda_k$ into a sum of two terms 

\beq
\vLambda_k = \he_k\cdot\nabla\vV(\vx) d +  \delta\vLambda_k \equiv \vDelta_k(\vx) + \delta\vLambda_k
\label{Lambdak}
\eeq
with $\he_k\equiv \vR_k/|\vR_k|$ a unit vector extending from the centroid of grain $i$ to the centroid of grain $j$ and $\vV(\vx)$ the coarse-grained velocity field. 
The first term on the right hand side of (\ref{Lambdak}) contains a random part - the direction of $\he_k$ - and a persistent part that is nothing but the strain rate tensor $T$. It should be noted that the first term does not average to zero because the average of $\he_k$ does not vanish since it always has a component pointing from the centre of the volume element $V_A$ to that of $V_B$. 
Expression (\ref{Lambdak}) assumes implicitly that averaging the grain rotations does not lead to a coarse-grained macroscopic rotation field, which is why the velocity field of the centres is the only coarse-grained quantity entering the description of the local velocity difference $\vLambda_k$. Thus, the relative velocities due to rotations of the grains at the points of contact are assumed to enter only as part of the fluctuation $\delta\vLambda_k$, whose average vanishes. 
This is why we use $\vDelta_k(\vx)$ in eq. (\ref{Lambdak}) rather than $\vLambda_k(\vx)$, as $\vDelta_k$ in eq. (\ref{kvelocity}) is the difference of centre of mass velocities.

Next, we separate the local normal force applied by particle $i$ to particle $j$ into an average and a fluctuation, $\vN_k = \vN(\vx) + \delta\vN_k$.
As the pair members move relative to one another the friction force that the $A$ member applies to the $B$ member is

\beq
\vF_k = \mu_d |\vN(\vx)+\delta\vN_k| \frac{\vLambda_k(\vx) + \delta\vLambda_k}{|\vLambda(\vx) + \delta\vLambda_k|}
%\vF_k = \mu_d |\bN(\vx)+\delta\bN_k|\frac{\vLambda_k(\vx) +  \delta\vLambda_k}{|\vLambda(\vx) + \delta\vLambda_k|}
\label{FrictionForce}
\eeq
To obtain the average friction force we need to average not only over the fluctuations $ \delta\vN$ and $ \delta\vLambda$, an average that will be denoted by $\langle ...\rangle$, but also over the directions of $\he_k$, which will be denoted by $\left[ ...\right]_{dir}$.
To average over the fluctuations requires expansion of the absolute values. However, the fluctuations need not be small and, wishing to avoid the assumption of small fluctuation, we use a different expansion, described in the following. Defining the average of the squared local contact force near $\vx$, $\Pi(\vx) = \langle \vN_k^2\rangle$, we expand $|\vN(\vx)+\delta\vN_k|$:

\beq
|\vN(\vx)+\delta\vN_k(\vx)| \approx \Pi(\vx)^{1/2}\left[ 1 + \frac{1}{2}\frac{|\vN_k|^2 - \Pi(\vx)}{\Pi(\vx)}\right]
\label{NormalForceExpansion}
\eeq
To arrive at a rotation covariant form for the stress, we need to consider next the invariance under rotations.  Since  $|\vN(\vx)|^2$ is a scalar, it is invariant under rotations and so is it average, $\Pi(\vx)$. Therefore, we can average $|\vN(\vx)|^2$ over the three mutually perpendicular planes, $\beta=1,2,3$, intersecting at a point near $\vx$, 
$\vN_k^2 = \frac{1}{3}\sum_{\beta=1}^3 \vN_{\beta k}^2$ and $\Pi(\vx)=\frac{1}{3}\sum_{\beta=1}^3 \left[\vN_{\beta}(\vx)^2+\langle\delta \vN_{\beta k}(\vx)^2\rangle\right]$. This is tantamount to averaging over three pairs of volumes, $V_A$ and $V_B$, oriented in orthogonal directions at $\vx$.
Similarly we define $\Gamma_k(\vx)=\frac{1}{3}\sum_{\beta=1}^3 \left[\vDelta_{\beta k}^2(\vx)+\langle\delta \vLambda_{\beta k}^2\rangle\right]$, where the index $k$ on $\Gamma$ reflects the dependence on the direction $\he_k$.
Similarly to (\ref{NormalForceExpansion}) we write 

\beq
|\vDelta_k(\vx)+\delta\vLambda_k(\vx)| \approx \Gamma(\vx)^{1/2}\left[ 1 + \frac{1}{2}\frac{|\vLambda_k|^2 - \Gamma(\vx)}{\Gamma(\vx)}\right]
\label{FrictionForceExpansion}
\eeq
where $\Lambda_k^2$ is also expressed in a rotation invariant form.

Next we use expressions (\ref{NormalForceExpansion}) and (\ref{FrictionForceExpansion}) to average the friction force given in equation (\ref{FrictionForce}), again avoiding expansion in the fluctuations $\delta\vN_k$ and $\delta\vLambda_k$, which need not be small. Again, we keep to first order, which is sufficient to bring out all the relevant physics. The average friction force, is then given by (using the convention of summation over repeated variables)

\begin{eqnarray}
& \vF_\beta(\vx) & = \mu_d\frac{\Pi(\vx)}{\Gamma(\vx)}
\bigg\{
\langle\vDelta_{\beta k}(\vx)\rangle +  \nonumber \\
& + & \frac{1}{3\Pi(\vx)}\big\{\left(\vN_\gamma(\vx)\cdot\delta\vN_{\gamma k}(\vx)\right)\delta\vLambda_{\beta k}(\vx) \nonumber \\
& + & \frac{1}{2}\langle\left(\delta\vN_{\gamma k} - \langle\delta N_{\gamma k}^2\rangle\right)\delta\vLambda_{\beta k}\rangle  - \nonumber \\
& - & \frac{1}{3\Gamma_k(\vx)}\big\{
\left(\vDelta_{\gamma k}(\vx)\cdot\delta\vLambda_{\gamma k}(\vx)\right)\delta\vN_{\beta}(\vx) - \nonumber \\
& - & \frac{1}{2}\langle\left(\delta\vLambda_{\gamma k}^2 - \langle\delta\vLambda_{\gamma k}^2\rangle\right)\delta N_{\beta}\rangle
\big\}
\bigg\}
\label{Fbeta}
\end{eqnarray}

To calculate these averages we need information about the correlations. 
The rubbing condition (\ref{RubCon}) implies a number of correlations that depend on the coarse-grained quantities. For example, its average is

\beq
\langle\delta\vN_{\beta k}\cdot\delta\vLambda_{\beta k}\rangle = -\vN_{\beta}(\vx)\cdot\vDelta_{\beta k}(\vx)
\label{RubConCorr1}
\eeq
No summation over $\beta$ implied in (\ref{RubConCorr1}), it holds for each $\beta$.
Another relation is obtained as follows. We first express the rubbing condition in the form

\beq
- \delta\vN_{k}\cdot\vDelta_{k}(\vx)  = \vN(\vx)\cdot  \vDelta_k(\vx) + \delta\vLambda_{k}\cdot\vN(\vx) + \delta\vN_{k}\cdot\delta\vLambda_{k} 
\label{RubConCorr2i}
\eeq
Squaring both sides of (\ref{RubConCorr2i}) and averaging over contacts inside $a$, assuming symmetry under reflections and rotations, gives 

\begin{eqnarray}
\left\langle\delta\vN_{k}\left( \delta\vN_{k}\cdot\delta\vLambda_{k}\right)\right\rangle & = & 0 \label{Symm1} \\ 
\left\langle\delta\vLambda_{k}\left( \delta\vLambda_{k}\cdot\delta\vN_{k}\right)\right\rangle & = & 0  \label{Symm2} \\
\left\langle\left( \delta\vN_{\alpha k} \cdot \delta\vDelta_{\alpha k}\right)^2\right\rangle & = & \frac{1}{3} \left[ \vDelta_{\beta k}^2\right]_{dir}\left\langle\delta\vN_{\gamma k}^2\right\rangle \label{Symm3} \\ 
\left\langle\left( \delta\vLambda_{\alpha k} \cdot \delta\vN_{\alpha}\right)^2\right\rangle & = & \frac{1}{3} \left[\vN_{\beta}^2\right]_{dir} \left\langle\delta\vLambda_{\gamma k}^2\right\rangle \label{Symm4}
\end{eqnarray}
Interchanging the roles of $(\vN,\delta\vN)$ and $(\vDelta,\delta\vLambda)$ and repeating the above analysis we obtain

\begin{eqnarray} 
\left\langle\delta\vN_{\alpha k}^2\right\rangle & = & \theta \vN_{\alpha}^2(\vx) \label{s3} \\
\left\langle\delta\vLambda_{\alpha k}^2\right\rangle & = & \theta \left[\vDelta_{\alpha}^2(\vx)\right]_{dir} \label{s4}
\end{eqnarray}
where $\theta$ is a dimensionless coefficient and a summation over $\alpha$ is implied.

From the definition $\vDelta_k(\vx)\equiv \he_k\cdot\nabla\vv(\vx)$ there follow the directional averages 

\begin{eqnarray}
\left[ \Delta^{(i)}_{\alpha k}(\vx)\right]_{dir} =  \frac{d}{2}\halpha_l T_{li}(\vx) \label{DirAve1} \\
\left[ \Delta^2_{\beta k}(\vx)\right]_{dir} =  d^2 |T(\vx)|^2 
\label{DirAve2}
\end{eqnarray}
where $\halpha_l$ is the $l$-component of a unit vector pointing from $V_A$ to $V_B$ perpendicular to the plane separating them, $i$ is a Cartesian component and $|T|=\left(T\cdot T\right)^{1/2}$ is the norm of $T$. 

Eqs. (\ref{Fbeta}), (\ref{RubConCorr1}) and (\ref{Symm3})-(\ref{DirAve2}) can now be used to obtain $\vF_{\alpha}(\vx)$. 
The force densities on the surface $\alpha$ at $\vx$, $\vnu_\alpha(\vx)$ and $\vphi_\alpha(\vx)$, are the three  $\alpha$ entries of the tensors $\sig^{(n)}(\vx)$ and $\sig^{(f)}(\vx)$, respectively. Thus, using eqs (\ref{NormalForceDensity}) and (\ref{FrictionForceDensity}) we obtain 

\beq
\sig^{(f)} = \mu_d \left[ \frac{ |\sig^{(n)}|}{|T|} \left( 1 + \frac{\theta}{6\left(1+\theta\right)} \right)T - 
\frac{1}{6\left(1+\theta\right)} \sigma^{(n)} \right]
\label{SigFricA}
\eeq
Two comments should be made about this result.
Firstly, in addition to the expected unit tensor $T/|T|$, this expression contains a term proportional to $\sigma^{(n)}$, seemingly providing friction even when the strain rate vanishes. This term poses no problem because the analysis is done only for plug-free regions, where the strain rate never vanishes.
Secondly, $\sig^{(f)}$ depends on the dimensionless parameter $\theta > 0$. It is tempting to conjecture that this parameter depends on the inertial number $I=|\dot{\gamma}|d/\sqrt{P/\rho_s}$, where $\dot{\gamma}$, $d$, $P$ and $\rho_s$ are, respectively, the shear rate, particle typical size, pressure and and particle specific mass\cite{daCetal05}. However, to attempt the derivation of this dependence is beyond the scope of this paper. This said, $\theta$ is small and is not expected to affect significantly the analysis.

To illustrate the use of this result, we apply it to an incompressible and isotropic fluid. 
Incompressibility implies a divergence-free velocity, leading to a traceless strain rate tensor, $Tr\{T\left(\vx \right)\}=0$. 
We assume that the friction coefficient is small. This allows us to use the normal strain tensor in the absence of friction for the symmetric part of normal stress tensor, which in turn is assumed to depend on the pressure alone, $\sigma_0^{(n)}(\vx)=P_0(\vx)\hat{\mathbb{1}}$, where $\hat{\mathbb{1}}$ is the unit tensor.
As we shall see below, the solid friction contributes to the stress a term that is a homogeneous function of degree zero in the strain rate. Our model aims at low strain rates, when this term is more significant than the viscosity term, which is linear in the strain rate. The latter would add to $\sigma_0^{(n)}$, but we ignore it here.

Using the subscripts $S$ and $A$ to denote, respectively, the symmetric and antisymmetric parts of a tensor, then in the presence of friction, the normal stress tensor is

\beq
\sig^{(n)}(\vx) = \sig_0^{(n)}(\vx) + \sig_A^{(n)}(\vx) 
\label{SigNormal}
\eeq
with $\sig_A^{(n)}$ balancing the antisymmetric part of the frictional stress tensor. It is important to note that the strain rate tensor may well contain an antisymmetric part, but the stress tensor must be symmetric to maintain torque balance on every volume element in the fluid. It follows that 

\beq
\sig_A^{(n)}(\vx) = -\mu_d\frac{1+7\theta}{6(1+\theta)}\frac{|\sig^{(n)}(\vx)|}{|T(\vx)|}T_A(\vx)  
\label{NormalStressAntisym}
\eeq
In terms of $\sig^{(n)}$, the total stress tensor is then

\beq
\sig(\vx) = \left(1 - \frac{\mu_d}{6(1+\theta)}\right)\! P_0\hat{\mathbb{1}} + 
\mu_d\! \left(1 + \frac{\theta}{6(1+\theta)}\right)\! \frac{|\sig^{(n)}|}{|T|}\! T_S  
\label{TotalStressA}
\eeq
Using $|\sig^{(n)}|=\sqrt{|\sig_S^{(n)}|^2+|\sig_A^{(n)}|^2}$ and eq. (\ref{NormalStressAntisym}), we obtain, to second order in $\mu_d$,

\beq
|\sig^{(n)}|=\sqrt{3}P_0\left\{1 + \frac{\mu_d^2}{2}\left[\frac{1+ 6\theta}{6\left(1 + \theta\right)} \right]^2 \frac{|T_A|^2}{|T|^2} \right\}
\label{SigNormalExplicit}
\eeq
It follows that the complete stress tensor to first order is

\beq
\sig(\vx) = \left(1 - \frac{\mu_d}{6(1+\theta)}\right)\! P_0\hat{\mathbb{1}} + 
\sqrt{3} \mu_d\! \left(1 + \frac{\theta}{6(1+\theta)}\right)\! P_0(\vx)\!\frac{T_S}{|T|}  
\label{TotalStressB}
\eeq

Tracing both sides of (\ref{TotalStressB}), using the incompressibility condition $Tr\{T\}=Tr\{T_S\}=0$ and that $P(\vx)=\left[1 - \frac{\mu_d}{6(1 + \theta})\right]P_0(\vx)$, we obtain the complete stress tensor

\beq
\sig(\vx) = P(\vx)\!\left[ \hat{\mathbb{1}} + \sqrt{3} \mu_d\! \left(1 + \frac{\theta}{6(1+\theta)}\right)\!\left(1 + \frac{\mu_d}{6(1+\theta)}\right) \frac{T_S}{|T|}\right]
\label{SigFull}
\eeq
This expression differs from the empirical one proposed in\cite{Sc87} and \cite{Jop06} in the intriguing dependence of the stress on the antisymmetric part of the strain rate.

To conclude, we have derived the stress tensor of plug-free flow of dense granular fluids from first principles in the regime where the viscosity contribution to the stress, which is linear in the strain rate, is negligible compared to that of solid friction, which is a homogeneous function of degree zero in the strain rate. 
A key result is the dependence of the stress tensor (\ref{SigFull}) on both the symmetric and the antisymmetric parts of the strain rate tensor. This dependence, missed in the existing empirical models\cite{Sc87,Jop06}, implies that local rigid-like rotation, which is a part of generic flow patterns, cannot be ignored and affects the  evolution of the flow. Another novel result is the explicit dependence of the stress tensor on the local interaction statistics through the parameter $\theta$.

This derivation combines with the description of plug formation and dynamics\cite{BlScEd08,ScBl11} to form a complete theory of dense granular flow - a phenomenon significant to many natural processes, technological applications and research disciplines.
In our view, further development of this model should include:
(a) construction of numerical flow codes, incorporating plug free flow as well as plug formation and dynamics;
(b) going beyond the approximations used here to improve the stress tensor;
(c) theoretical studies of flows that can be tested against amenable experiments.

\end{document}